\documentclass[twocolumn,preprintnumbers,amsmath,amssymb,superscriptaddress]{revtex4}
\usepackage{amsmath}
\usepackage{amsfonts}
\usepackage{amssymb}
\usepackage{graphicx}
\usepackage{bm}
\usepackage{mathenv}
\usepackage{color}

\begin{document}

\preprint{APS/PRB}

\title{The magnetic phase of the perovskite CaCrO$_3$ studied with $\mu^{+}$SR}

\author{Oren~Ofer}
 \email{oren@triumf.ca}
\affiliation{TRIUMF, 4004 Wesbrook Mall, Vancouver, BC, Canada V6T 2A3}
\author{Jun~Sugiyama}
\affiliation{Toyota Central Research and Development Labs. Inc., Nagakute, Aichi 480-1192, Japan}
\author{Martin~M\aa{}nsson}
\affiliation{Laboratory for Neutron Scattering, ETH Z\"{u}rich and Paul Scherrer Institute, CH-5232 Villigen PSI, Switzerland}
\author{Kim~H.~Chow}
\affiliation{Department of Physics, University of Alberta, Edmonton, AB, Canada, T6G 2G7}
\author{Eduardo~J.~Ansaldo}
\affiliation{TRIUMF, 4004 Wesbrook Mall, Vancouver, BC, Canada, V6T 2A3}
\author{Jess~H.~Brewer}
\affiliation{TRIUMF, 4004 Wesbrook Mall, Vancouver, BC, Canada, V6T 2A3}
\affiliation{CIfAR and Department of Physics and Astronomy, University of British Columbia, Vancouver, BC, Canada, V6T 1Z1}
\author{Masahiko~Isobe}
\author{Yutaka~Ueda}
\affiliation{Materials Design and Characterization Laboratory, Institute for Solid State Physics, University of Tokyo, Kashiwa, Chiba 277-8581, Japan}

\date{\today}

\begin{abstract}

We investigated the magnetic phase of the perovskite CaCrO$_3$ by using the muon spin relaxation technique accompanied by susceptibility measurements. A thermal hysteresis loop is identified with a width of about 1~K at the transition temperature.  Within the time scale of the muon lifetime, a static antiferromagnetic order is revealed with distinct multiple internal fields which are experienced in the muon interstitial sites below the phase-transition temperature, $T_N=90~K$. Above $T_N$, lattice deformations are indicated by transverse-field muon-spin rotation and relaxation suggesting a magneto-elastic mechanism.
\end{abstract}



\maketitle

\section{\label{sec:I}Introduction}

The co-existence of metallic conductivity \textit{and} 
antiferromagnetism is uncommon and has therefore
inspired theoretical \cite{streltsov} and experimental \cite{zhou,Komarek} interests,
particularly for transition metal oxides ($M$O$_x$),  due to a strong hybridization of $d$ orbital of the metal ion ($M$) and 2$p$ orbital of the oxygen (O).
The magnetic properties in these systems are governed
mainly by a super-exchange interaction $via$ an oxygen
between the nearest neighboring (NN) $M$ ions,
and additionally by a competition
between the NN interaction
and the next-nearest-neighbor interaction.
This competition has been reported to cause the magnetic order
and affect the structural and electronic properties.
In most experimental systems \cite{Yoshida,Jun_CaNaVO},
these interactions are characterized by low dimensionality
thereby granting the exotic coexistence of the antiferromagnetic (AFM) order and metallic conductivity. The three-dimensional perovskite CaCrO$_3$ represent an exception.


Although CaCrO$_3$ was reported to show semiconducting \cite{goodenough}
or insulating \cite{zhou} properties with AFM order below 90~K (=$T_{\rm N}$),
recent measurements on single crystals,  grown in high-pressure synthesis,
indicated metallic conductivity for CaCrO$_3$ even below $T_{\rm N}$\cite{weiher}. This was also supported by infrared reflectivity measurements on a polycrystalline sample\cite{Komarek}. Furthermore, based on powder neutron diffraction analysis,
the AFM spin structure was proposed as a $C$-type AFM,
in which Cr spins order antiferromagnetically in the $ab$ plane,
but ferromagnetically along the $c$ axis \cite{Komarek}.
A high-temperature Curie-Weiss fit of the susceptibility-versus-$T$ curve in the $T$ range between 400 and 600~K indicated the effective magnetic moment ($\mu_{\rm eff}$) of Cr to be 3.6~$\mu_{\rm B}$
and the Weiss $T$ ($\Theta_{\rm p}$) is about -920~K\cite{weiher,Castillo}.
Since the Cr ions are in a ${4+}$ state with $S$=1, the obtained $\mu_{\rm eff}$ is rather high compared with the localized-spin-only value (2.85~$\mu_{\rm B}$).
Moreover, it was found that the CaCrO$_3$ system has
a magneto-elastic distortion driven by the Cr moments \cite{Komarek},
resulting in a small change in the orthorhombic lattice parameters at $T_{\rm N}$.
Despite the several macroscopic measurements mentioned above,
the microscopic magnetic nature of CaCrO$_3$ has not been investigated so far,
because of the difficulty of sample preparation and
eventually the absence of NMR-active elements in CaCrO$_3$,
as the natural abundance for the NMR-active $^{43}$Ca is 0.14$\%$
and that for $^{53}$Cr is 9.5$\%$.

\begin{figure}[t]
  \begin{center}
    \includegraphics[keepaspectratio=true,width=60 mm]{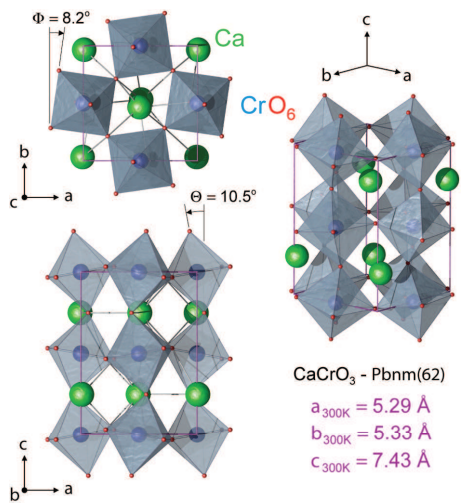}
  \end{center}
  \caption{(Color online)
The GdFeO$_{3}$-type crystal structure (space group $Pbnm$, No. 62) of CaCrO$_{3}$ that evolves from the ideal perovskite structure by a rotation ($\Phi\approx$~8.2$^{\circ}$) and tilt ($\Theta\approx$~10.5$^{\circ}$) of the CrO$_{6}$ octahedra \cite{Komarek}. Thin purple (light grey) lines enclose the unit cell.}
  \label{fig:structure}
\end{figure}

In contrast to NMR, the
 muon-spin rotation and relaxation ($\mu^+$SR) technique
is applicable to all magnetic materials \cite{AmatoMuSR,brewerency},
even if they lack elements with nuclear magnetic moments.
We have, therefore, carried out an experimental study
on the perovskite CaCrO$_3$  (see Fig.~\ref{fig:structure})
by means of $\mu^+$SR due to its remarkable ability in detecting 
local magnetic order,
whether it is short- or long-ranged. Combining bulk dc-susceptibility and $\mu^+$SR measurements, we characterize the magnetic properties of CaCrO$_3$.
Our major finding is a C$_\text{y}$-type AFM phase, determined from the multiple frequencies revealed in zero-field (ZF) $\mu^+$SR. Such frequencies are the attributes of the several $\mu^+$ interstitial sites in the AFM phase. Second, the transverse-field (TF) $\mu^+$SR signals also suggests a deformation of the lattice, thereby corroborating that CaCrO$_3$ experiences a magneto-elastic distortion.

\section{\label{sec:E}Experiment}
Polycrystalline CaCrO$_{3}$ was prepared by a solid state reaction of CaO and CrO$_{2}$ under 4 GPa at 1000$^{\circ}$C for 30 min in the Institute of Solid State Physics of University of Tokyo. dc-$\chi$ measurement (shown later) and powder diffraction were performed and agreed with previously published data. In the $\mu^+$SR experiment, the powder sample was placed in a small envelope made of very thin Al-coated Mylar tape and attached to a low-background sample holder. In order to make certain that the muon stopped primarily inside the sample, we ensured that the side facing the muon beamline was only covered by a single layer of the mylar tape. Subsequently, ZF,  TF and longitudinal-field (LF)  $\mu^+$SR spectra were collected for 1.8~K~$\leq{}T\leq$~150~K, at the {\bf M20} and {\bf M15} surface muon channel at TRIUMF, Vancouver, Canada. The experimental setup and techniques are described in detail elsewhere \cite{Kalvius}.

\section{\label{sec:R}Results}
\begin{figure}[t]
  \begin{center}
    \includegraphics[keepaspectratio=true,width=\columnwidth]{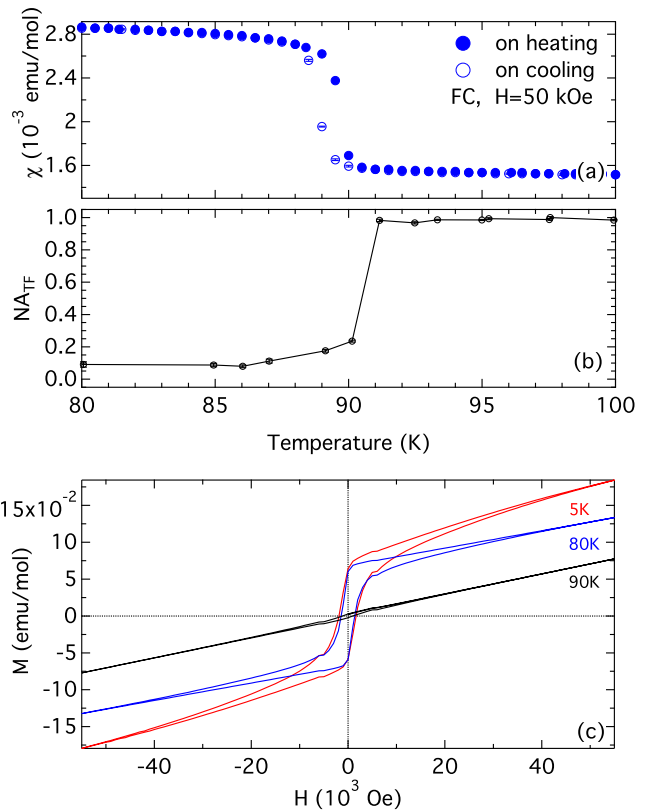}
  \end{center}
  \caption{(Color online)
  (a) The $\chi(T)$ curve at the vicinity of magnetic transition
  measured on cooling and on heating with $H=50~$kOe in FC mode.
  (b) The normalized wTF Asymmetry with applied field $H=30$~Oe.
  (c) The magnetization ($M$) versus $H$ at 90 (black line), 80 (blue), and 5~K (red).
  }
  \label{fig:chi}
\end{figure}

The bulk magnetization and susceptibility of CaCrO$_3$ is shown in Fig.~\ref{fig:chi} as a function of $T$ and $H$. The temperature dependence of the bulk dc susceptiblity, $\chi$, was measured using a commercial superconducting quantum interference device magnetometer (Magnetic Property Measurement System - Quantum Design). The magnetic transition is seen clearly at $T_N\sim90$~K, however the $\chi(T)$ curve shows the appearance of small spontaneous magnetization below $T_{\rm N}$, that is not of a typical AFM transition. The same $\chi(T)$ behavior is seen at all fields measured ($1$~kOe$\leq H\leq50$~kOe, not shown), therefore this can be ruled-out as field-induced. Figure \ref{fig:chi}(a) displays the characteristic $\chi(T)$ curves obtained on cooling and heating with an applied field of 50~kOe.   In Fig.~\ref{fig:chi}(c) we plot the magnetization, $M$, versus the field, $H$, taken at several temperatures,  the magnetic hysteresis loop seen below $T_N$, at 80 and 5~K, indicates the presence of a weak ferromagnetic contribution, which is attributed from the canted spins along the $c$ direction. To corroborate this magnetic transition, we performed weak transverse-field measurements.

A weak transverse-field (wTF) measurements, where the field applied is perpendicular to the muon spin direction and is weak compared to any spontaneous internal fields in the ordered phase, is a sensitive probe to local magnetic order.  In Fig.~\ref{fig:chi}~(b), we plot the $T$ dependence of the normalized wTF asymmetry  ($NA_{\text TF}$) taken at $30$~Oe on heating. Note that the $NA_{\text TF}$ is proportional to the volume fraction of the paramagnetic phase. The magnetic phase transition, as indicated by the wTF measurements, also identifies the transition at $T_N=90$~K.  To further explore the magnetic phase we performed ZF $\mu^+$SR, which is a site-sensitive probe in which the muon asymmetry is only affected by the internal magnetic fields.
\begin{figure}[tbp]
\begin{center}
\includegraphics[width=\columnwidth]{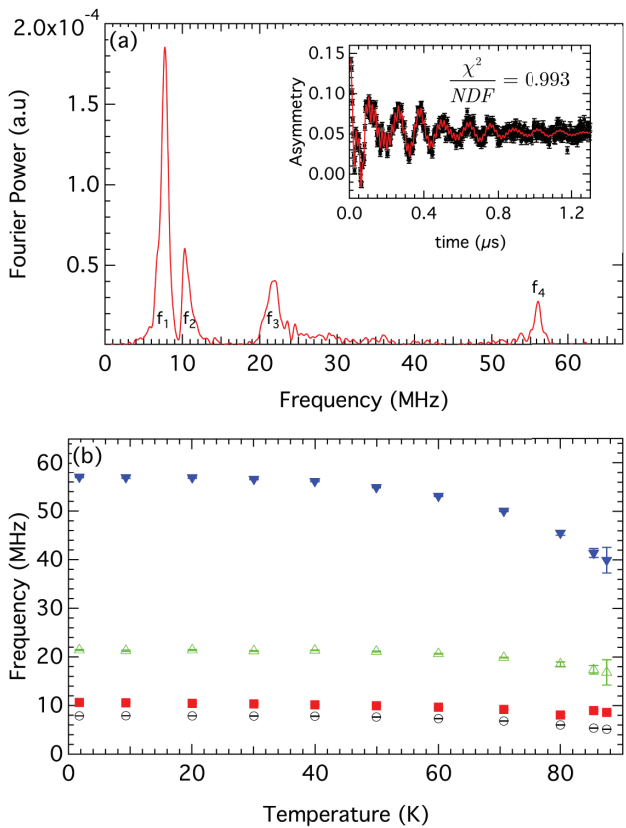}
\end{center}
\caption{(Color online) (a) The Fourier Transform of the Zero Field spectra at $T=40~K$. Inset shows the raw ZF Asymmetry spectrum. (b) The temperature dependence of the muon precession frequencies at the AFM phase. }
\label{fig:ZF}
\end{figure}

ZF $\mu^+$SR measurements were taken in the same configuration as described in Sec. \ref{sec:E}. The inset of Fig.~\ref{fig:ZF}(a) shows the raw data taken at $T=40$~K where the main figure displays the Fourier transform of the raw ZF spectrum.  At all temperatures below $T_N$, the ZF spectrum exhibits multiple frequencies, which consists of mainly four frequencies. Therefore, the spectra were fitted with a sum of four oscillating signals where the asymmetry is divided into a 2/3 oscillatory and 1/3 "tail" exponentially relaxing components, namely,
\begin{eqnarray*}
A_0 P_{\text ZF}(t)=&\sum_{i=1}^4 A_{\text{osc,}i} \cos(\omega_i t)\exp(-\lambda_{\text {osc},i}t)\nonumber\\&+A_{\text{tail,}i}\exp(-\lambda_{\text{tail,}i} t)
\end{eqnarray*}
where $A_\text{osc}=2A_0/3$ and $A_\text{tail}=A_0/3$  are the asymmetries, $\lambda_\text{osc}=1/T_2$ and $\lambda_\text{tail}=1/T_1$ are the relaxation rates and $\omega_i$ are the muon Larmor frequencies of the four signals. The quality of the fit is demonstrated in the inset of Fig.~\ref{fig:ZF}(a) by the solid line.

Figure \ref{fig:ZF}(b) summarizes the $T$ dependence of the muon precession frequencies ($f_i=\omega_i/(2\pi)$). The low frequencies, $i=1,2,3$~, are nearly independent of temperature up to $90$~K, whereas $f_4$ decreases smoothly above $30$~K. The amplitude (not shown) of $f_4$ also decreases dramatically above $50$~K.  We, therefore, suggests that CaCrO$_3$ displays a single AFM phase. Electrostatic potential calculations confirm that there are four possible different muon sites near O$^{-2}$ ions in the lattice.Assuming the proposed C$_\text{y}$-AFM spin structure \cite{Alario,Komarek}, we use dipolar field calculation to describe these internal magnetic fields. Since the amplitude of $f_4$ is small and difficult to accurately estimate throughout the whole temperature range (amplitude diminishes above 50~K and there is considerable variation in the range $2\leq T\leq50$~K), we disregard $f_4$ in the calculation and aim at characterizing the other three frequencies only. The calculated internal fields ($H_{\text{int},i}$ where $i=1,2,3,4$) are proportional to the ordered moment $\mu_\text{Cr}$ and obey $f_i=H_{\text{int},i}\times13.554$(kHz/Oe).  We find $H_{\text{int},i}/\mu_\text{Cr}$=2.5(2),~3.16(7)~kOe/$\mu_B$, and 8.7(2)~kOe/$\mu_B$ for $i=1-3$ respectively.  We find that the ratio between the fields $H_{\text{int},1}/H_{\text{int},3}=0.28(9)$ and $H_{\text{int},2}/H_{\text{int},3}=0.36(7)$.  The observed $f_1/f_3=0.367(1)$ and $f_2/f_3=0.495(2)$, thus the overall calculated result is within $\sim25\%$ error. Furthermore, the absolute values of $f_i$ allows us to estimate that the ordered moment $\mu_\text{Cr}(1.8~\text{K})=1.38(22)\mu_{\rm B}$. This value agrees well with the estimated $\mu_\text{ord}=1.2\mu_{\rm B}$ from recent neutron scattering. The same calculation for a C$_\text{x}$-AFM structure yields that $H_{\text{int},1}/H_{\text{int},3}=0.29(3)$, $H_{\text{int},2}/H_{\text{int},3}=0.88(7)$, and $\mu_\text{Cr}(1.8~\text{K})=1.2(4)~\mu_{\rm B}$. Although the C$_\text{y}$-AFM $H_\text{int}$ ratios seem to agree well between the calculation and the experiment, $\mu_\text{Cr}$ is slightly higher than the proposed C$_\text{y}$-type but within the error. Nevertheless, it is difficult to determine which AFM structure is more reasonable for the ground state of CaCrO$_3$ based only on the present $\mu^+$SR results. The comparison between $\mu_\text{ord}$ estimated by neutron and that by $\mu^+$SR is most likely to support a C$_\text{y}$-AFM structure, as proposed by neutron measurements\cite{Komarek}.

\begin{figure}
\begin{center}
\includegraphics[width=\columnwidth]{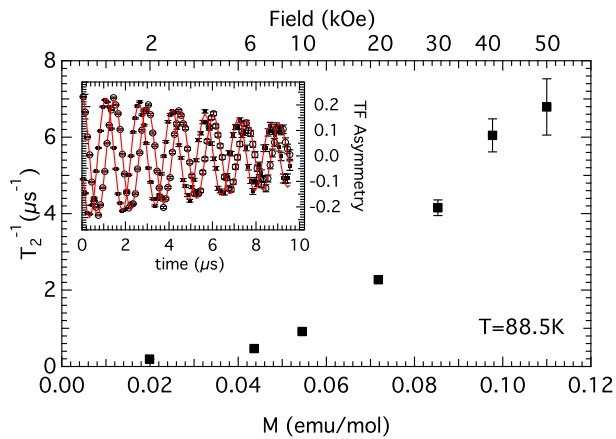}
\end{center}
\caption{(Color online) The TF Relaxation rate, $R_\text{TF}$ versus the magnetization, $M$. Inset, the muon decay asymmetry in Transverse Field of 2~kOe, rotated in a reference frame of $26.5$~MHz.}
\label{fig:RlxAsy}
\end{figure}

In order to study the magneto-elastic coupling, TF $\mu^+$SR measurements were taken at fields ranging from 2 to 50~kOe at the transition temperature. The inset of Fig.~\ref{fig:RlxAsy} plots the muon decay asymmetry in a transverse field of 2~kOe rotated at a reference frame\cite{tanyaRRF} of $26.5$~MHz. Since a high magnetic field is available only along the axial direction, TF measurements were performed in a spin-rotated mode, where the muon spin is rotated 90$^\circ$ thereby perpendicular to the field direction. The LF measurements, were performed in a non-spin-rotated mode, therefore the spin and the field were parallel. No temperature dependence was found on the muon relaxation in the LF measurements, and was at least an order of magnitude smaller than the TF relaxation. It should be noted that TF measurements probe both the static and dynamic field fluctuations and the LF measurements probes the dynamic fluctuations only. Hence, it was found that the dynamic fluctuations are weak compared to the static fluctuations, therefore negligible.

The TF muon-spin polarization is best described by
\begin{equation}
A_0P_\text{TF}=A_0\exp(-R_\text{TF}t)\cos(\omega t+\varphi)
\end{equation}
where $R_\text{TF}=(T_2^\ast)^{-1}$ is the TF relaxation rate and $\omega~=~\gamma_\mu~H_\text{TF}$.  The quality of the fit is described by the solid line in the plot. Figure ~\ref{fig:RlxAsy} depicts $R_\text{TF}$ vs the bulk magnetization. $R_\text{TF}$ increases slowly with increasing magnetization, $M$. At $M\sim0.06$~emu/mol, $R_\text{TF}$ increases more rapidly with $M$. In order to explain this occurrence, one should observe the muon behavior under external fields.

The field at the muon site is given by $\textbf{B}=(1+{\bf A}\chi){\bf H}_\text{TF}$, where ${\bf A}$ is the coupling between the muon and the neighboring electron. We assume that the coupling depends on the muon-electron distance and is isotropic, therefore we write $A$ as sum of a mean $\overline{A}$ and a fluctuating $\delta A$ components. Thus one finds for a gaussian distribution of couplings, $\rho(A)=1/\sqrt{2\pi\sigma}\cdot\exp(-(A/\sqrt{2}\sigma)^2)$,  the spectrum width obeys\cite{CarretaKeren},
\begin{equation}
R_\text{TF}=\gamma_\mu M\cdot\sigma.
\label{eq:dist}
\end{equation}
If $\overline A$ and $\sigma$ are temperature independent, $R_\text{TF}$ is expected to be linearly proportional to $M$. Figure~\ref{fig:RlxAsy} indicates that $R_\text{TF}$ is not proportional or does not depend linearly on $M$. This suggests a modification in the hyperfine coupling, $A$, which is expected when lattice distortions take place\cite{CarretaKeren}. In fact, by calculating the temperature dependence (for $90\leq T\leq 130~$K, not shown) of $\sigma$, using Eq.~(\ref{eq:dist}), we can calculate $\Delta\sigma/\overline{\sigma}$, where $\Delta\sigma$ and $\overline{\sigma}$ are the standard deviation and average of $\sigma$ respectively. The ratio $\Delta\sigma/\overline{\sigma}$ gives a measure of the relative change in the variation of the muon-electron distance due to temperature changes. We find that $\Delta\sigma/\overline{\sigma}$ is $\sim208\%$, for comparison, compounds which do not distort have a much smaller $\Delta\sigma/\overline{\sigma}$, for example, the pyrochlore Tb$_2$Ti$_2$O$_7$ has $\Delta\sigma/\overline{\sigma}$ of $15\%$ \cite{orenTbTiO}.   A possible reason for such a distortion is a response of the lattice to the magnetic interactions through a magneto-elastic coupling \cite{eva}.  Such coupling relieves the frustrated Cr interactions by causing lattice deformations.  This is consistent with the scenario suggested by neutron scattering studies \cite{Komarek}.

\section{\label{sec:S}Summary}
By means of $\mu^+$SR and susceptibility, we clarified the nature of the magnetic phase of the perovskite CaCrO$_3$ below $T_N$. Susceptibility measurements demonstrate a  presence of a weak ferromagnetic contribution in the magnetic hysteresis loops. However, $\mu^+$SR demonstrates the formation of static AFM order below $T_{\rm N}$. Based on electrostatic and dipole field calculations, the C$_\text{y}$-AFM structure, which was proposed by neutron measurements, 
is found to be  the most reasonable in explaining the internal magnetic fields detected by zero-field $\mu^+$SR measurements. 

Transverse-field $\mu^+$SR measurements in the paramagnetic phase provided 
the temperature dependence of the spin-spin relaxation rate ($T_2^{-1}$). 
The lack of a linear relationship between $T_2^{-1}$ and magnetization above the vicinity of $T_{\rm N}$ 
suggests the presence of magneto-elastic coupling, which induces the abrupt change in the lattice parameters at $T_{\rm N}$, and reduce the frustrated interactions.

Although the overall $\mu^+$SR results confirm the past neutron results, 
it should be noted that the time window and spatial resolution of $\mu^+$SR are different from 
those of neutron scattering. Therefore, combining the past neutron results, 
the magnetic and structural nature of CaCrO$_3$ have been fully elucidated by this work.

\begin{acknowledgments}
We are grateful to the staff of TRIUMF for assistance with the $\mu^+$SR experiments.
JHB is supported at UBC by CIfAR, NSERC of Canada, and (through TRIUMF) by NRC of Canada and
KHC by NSERC of Canada and (through TRIUMF) by NRC of Canada.
This work is also supported by Grant-in-Aid for Scientific Research (B), 19340107, MEXT, Japan.
\end{acknowledgments}


\end{document}